\documentclass[lettersize,journal]{IEEEtran}
\usepackage{amsmath,amsfonts}
\usepackage{algorithmic}
\usepackage{algorithm}
\usepackage{array}
\usepackage[caption=false,font=normalsize,labelfont=sf,textfont=sf]{subfig}
\usepackage{textcomp}
\usepackage{stfloats}
\usepackage{url}
\usepackage{verbatim}
\usepackage{graphicx}
\usepackage{cite}
\usepackage{tabularx}
\usepackage{booktabs} 
\begin{document}

\title{Improving multidimensional projection quality with user-specific metrics and optimal scaling}

\author{Maniru Ibrahim\\

\thanks{M. Ibrahim is with Mathematics Applications Consortium for Science and Industry (MACSI), Department of Mathematics and Statistics,
University of Limerick, Limerick V94 T9PX, Ireland}
}

\markboth{}%
{}

\IEEEpubid{}

\maketitle

\begin{abstract}
The growing prevalence of high-dimensional data has fostered the development of multidimensional projection (MP) techniques, such as t-SNE, UMAP, and LAMP, for data visualization and exploration. However, conventional MP methods typically employ generic quality metrics, neglecting individual user preferences. This study proposes a new framework that tailors MP techniques based on user-specific quality criteria, enhancing projection interpretability.

Our approach combines three visual quality metrics, stress, neighborhood preservation, and silhouette score, to create a composite metric for a precise MP evaluation. We then optimize the projection scale by maximizing the composite metric value. We conducted an experiment involving two users with different projection preferences, generating projections using t-SNE, UMAP, and LAMP. Users rate projections according to their criteria, producing two training sets. We derive optimal weights for each set and apply them to other datasets to determine the best projections per user.

Our findings demonstrate that personalized projections effectively capture user preferences, fostering better data exploration and enabling more informed decision-making. This user-centric approach promotes advancements in multidimensional projection techniques that accommodate diverse user preferences and enhance interpretability.
\end{abstract}


\section{Introduction}
\IEEEPARstart{T}{he} rapid growth of high-dimensional data in diverse fields such as finance, biology, social sciences, and engineering has necessitated the development of effective data visualization and exploration techniques. Multidimensional projection (MP) techniques have emerged as powerful tools for representing high-dimensional data in lower-dimensional spaces, typically two or three dimensions, in a way that preserves the structure and relationships present in the original data \cite{maaten2008visualizing, mcinnes2018umap,Joia2011LocalAM, paulovich2008lap}. Examples of widely-used MP techniques include t-distributed Stochastic Neighbor Embedding (t-SNE) \cite{maaten2008visualizing}, Uniform Manifold Approximation and Projection (UMAP) \cite{mcinnes2018umap}, and Local Affine Multidimensional Projection (LAMP) \cite{paulovich2008lap}.

Despite the rapid advancements in MP techniques, the evaluation of projection quality remains a challenge. Many MP algorithms rely on generic quality metrics, such as stress, neighborhood preservation, and silhouette score, to evaluate the effectiveness of projections in preserving the data structure \cite{Lee2009QualityAO, pezzotti2017approximated}. However, these quality metrics do not always align with the perception and preferences of individual users, who are ultimately responsible for interpreting and analyzing the visualizations \cite{sedlmair2012taxonomy, bertini2011quality}. As a result, there is a growing need to develop personalized projection techniques that can accommodate diverse user preferences and facilitate more effective data exploration and decision-making.
Existing MP methods often employ generic quality metrics for evaluating projection quality, which may not capture the diverse preferences and criteria of individual users. Moreover, these metrics may not always correspond to the perceptual properties and interpretability of the resulting projections \cite{sedlmair2012taxonomy}. As a consequence, users may struggle to make sense of the visualizations and identify meaningful patterns or relationships in the data.

In order to address these challenges, it is crucial to develop a user-centric approach to MP that tailors the projection techniques based on user-specific quality criteria. Such an approach should not only enhance the interpretability of the resulting projections but also empower users to make more informed decisions and facilitate better data exploration.

This paper propose a new framework for personalized multidimensional projections that takes into account user-specific quality criteria to optimize projection quality and interpretability. The contributions are as follows:

\begin{enumerate}
\item We introduce a composite quality metric that combines three known visual quality metrics—stress, neighborhood preservation, and silhouette score—to provide a more precise assessment of MP quality. This composite metric allows for a more holistic evaluation of the projections and serves as a basis for adapting the MP techniques according to user preferences.
\item We propose an optimization approach for finding the best scale for a projection by maximizing the value of the composite quality metric. This optimization strategy ensures that the resulting projections are not only of high quality but also tailored to the user's specific criteria.

\item We demonstrate the effectiveness of our personalized projection framework through an experiment involving two users with distinct projection preferences. We generate various projections using t-SNE, UMAP, and LAMP by varying algorithm-specific parameters, such as perplexity and n\_neighbors. The users rate different projections according to their personal criteria, resulting in two distinct training sets. We learn the optimal set of weights for each training set and apply these weights to other datasets to find the best projections for each user. Our results show that the personalized projections are coherent with each user's criteria and effectively capture their preferences, facilitating better data exploration and decision-making
\end{enumerate}

This paper is organized as follows: Section 2 provides an overview of related work, including a brief introduction to the MP techniques used in this study (t-SNE, UMAP, and LAMP) and the visual quality metrics employed (stress, neighborhood preservation, and silhouette score). Section 3 presents the personalized projection framework, detailing the composite quality metric, the optimization approach for projection scaling, and the methodology for user-specific weighting of quality metrics. Section 4 outlines the experiment design, including the datasets used, the generation of projections, the user rating process, and the learning of optimal weights. Section 5 presents the results and evaluation of the experiment, comparing the personalized projections with traditional quality metrics and discussing the consistency of user preferences. Section 6 discusses the advantages and limitations of our approach, as well as the implications for data exploration and decision-making. Finally, Section 7 concludes the paper and suggests future research directions.

\subsection{Multidimensional projection techniques}

Multidimensional projection techniques are essential tools for visualizing high-dimensional data in lower-dimensional spaces, typically two or three dimensions. These techniques aim to preserve the structure and relationships present in the original data to facilitate effective data exploration and interpretation. In this section, we provide an overview of three widely-used multidimensional projection techniques: t-distributed Stochastic Neighbor Embedding (t-SNE), Uniform Manifold Approximation and Projection (UMAP), and Local Affine Multidimensional Projection (LAMP).

\subsubsection{t-SNE}

t-distributed Stochastic Neighbor Embedding (t-SNE) is a popular non-linear dimensionality reduction technique introduced by van der Maaten and Hinton \cite{maaten2008visualizing}. t-SNE aims to minimize the divergence between two probability distributions, one that captures pairwise similarities in the high-dimensional space and another that represents similarities in the low-dimensional space.

t-SNE constructs a probability distribution over pairs of high-dimensional datapoints in such a way that similar points have a high probability of being assigned to the same cluster, while dissimilar points have a low probability. It then constructs a similar probability distribution in the low-dimensional space and minimizes the Kullback-Leibler (KL) divergence between the two distributions using gradient descent.

One of the key parameters of t-SNE is perplexity, which balances the local and global aspects of the data structure. Lower values of perplexity emphasize local structures, while higher values give more weight to global structures. The choice of perplexity can significantly affect the resulting projection, and it is often recommended to explore multiple perplexity values to find the most appropriate projection for a given dataset \cite{maaten2008visualizing}.

\subsubsection{UMAP}

Uniform Manifold Approximation and Projection (UMAP) is another non-linear dimensionality reduction technique, proposed by McInnes et al. \cite{mcinnes2018umap}. UMAP is based on manifold learning and topological data analysis, and it seeks to preserve both the global and local structure of the data. UMAP can be considered a generalization of t-SNE, as it also aims to minimize the divergence between probability distributions in high-dimensional and low-dimensional spaces. However, UMAP introduces several innovations that improve its scalability, computational efficiency, and the preservation of global structure.

UMAP uses a fuzzy simplicial set representation to capture the topology of the high-dimensional data, and then optimizes an objective function that encourages the low-dimensional representation to have a similar topological structure. One of the main parameters of UMAP is $n-$neighbors, which determines the size of the local neighborhood considered when constructing the fuzzy simplicial set representation. Varying the $n-$neighbors parameter can impact the balance between local and global structure preservation in the resulting projection.

\subsubsection{LAMP}

Local Affine Multidimensional Projection (LAMP) is a linear projection technique introduced by Joia et al. \cite{Joia2011LocalAM}. LAMP aims to provide a fast, high-precision multidimensional projection by minimizing the reconstruction error of the original data points in the low-dimensional space. LAMP is based on the idea that local regions in high-dimensional space can be well-approximated by linear transformations, and it constructs a set of local linear projections by solving a least squares problem.

In LAMP, the user defines a set of control points \cite{paulovich2008lap} in the high-dimensional space and their corresponding coordinates in the low-dimensional space. LAMP then computes a local linear transformation for each data point using the control points and their weights. The weights are determined by a Gaussian function that depends on the distance between the data point and the control points. The choice of control points and their relative importance can significantly impact the resulting projection, and it is often necessary to explore different configurations to find the most suitable projection for a given dataset \cite{Joia2011LocalAM}.

\subsection{Visual quality metrics}

Visual quality metrics are essential for evaluating the effectiveness of multidimensional projection techniques, as they provide quantitative measures of how well the projections preserve the structure and relationships present in the original high-dimensional data. In this section, we discuss three widely-used visual quality metrics: stress, neighborhood preservation, and silhouette score.

\subsubsection{Stress}

Stress is a widely-used quality metric for assessing the preservation of pairwise distances in multidimensional projections \cite{Lee2009QualityAO,Joia2011LocalAM,Nonato2019MultidimensionalPF}. Stress measures the difference between the pairwise distances in the original high-dimensional space and the distances in the low-dimensional projection. Lower stress values indicate that the projection better preserves the original distances.

Stress is defined as follows:

\[ \textit{Stress}(P) = \sqrt{\frac{\sum_{i,j}(d_{ij} - d'{ij})^2}{\sum{i,j}d_{ij}^2}} \]

where $P$ is the projection, $d_{ij}$ is the distance between points $i$ and $j$ in the high-dimensional space, and $d'_{ij}$ is the distance between the same points in the low-dimensional projection.

While stress provides a useful measure of global structure preservation, it may not adequately capture the preservation of local neighborhood structures, which are often of greater interest in visualization tasks.

\subsubsection{Neighborhood preservation}

Neighborhood preservation (NP) is a quality metric that specifically focuses on the preservation of local neighborhood structures in multidimensional projections \cite{Lee2009QualityAO,Joia2011LocalAM,Nonato2019MultidimensionalPF}. Neighborhood preservation evaluates the extent to which the nearest neighbors of a point in the high-dimensional space are also its nearest neighbors in the low-dimensional projection.

Neighborhood preservation can be computed using various approaches, such as rank-based measures, set-based measures, or graph-based measures. A common method for quantifying neighborhood preservation is the rank-order correlation between distances in high-dimensional space and low-dimensional projection. High rank-order correlation values indicate that the projection better preserves the local neighborhood structures.

\subsubsection{Silhouette coefficient}

Silhouette coefficient (SC) is a quality metric that measures the clustering quality of a projection \cite{pezzotti2017approximated,Joia2011LocalAM}. The silhouette coefficient evaluates the degree to which points within the same cluster are closer to each other than to points in other clusters. A high silhouette score indicates that the projection better separates the clusters in the data.

Silhouette coefficient is defined as follows:

\begin{align*}
 s(i) &= \frac{b(i) - a(i)}{\max{(a(i), b(i))}} 
\end{align*}

where $s(i)$ is the silhouette score for point $i$, $a(i)$ is the average distance between point $i$ and all other points within the same cluster, and $b(i)$ is the average distance between point $i$ and all points in the nearest neighboring cluster.

The overall silhouette score for a projection is calculated as the average silhouette score across all points. It is important to note that the silhouette score assumes the presence of well-defined clusters in the data, and it may not be applicable to datasets with more complex, nonclustered structures.

\subsection{Challenges in evaluating multidimensional projection techniques}

Multidimensional projection techniques, such as t-distributed Stochastic Neighbor Embedding (t-SNE), Uniform Manifold Approximation and Projection (UMAP), and Local Affine Multidimensional Projection (LAMP), have become popular tools for visualizing high-dimensional data in lower-dimensional spaces. These techniques aim to preserve the structure and relationships present in the original data, facilitating effective data exploration and interpretation \cite{maaten2008visualizing, mcinnes2018umap, Joia2011LocalAM}. However, multidimensional projections are inherently vulnerable to errors and distortions due to the limitations of orthogonal mappings from multidimensional spaces to visual spaces \cite{sacha}. Consequently, the selection of an appropriate projection technique and its parameters is a critical step in the visualization process.

One of the main challenges in evaluating and selecting projections is the reliance on distance preservation-based metrics, such as stress and distance plots, which may not fully capture the preservation of local neighborhood structures that are often of greater interest in visualization tasks \cite{Mart}. Moreover, many existing quality metrics, such as the correlation coefficient, Kruskal's stress function, silhouette coefficient, and neighborhood preservation, focus on either global or local aspects of the data structure, but not both \cite{corr, stress, tan, Joia2011LocalAM}. The development of metrics that can assess multiple aspects of the data structure simultaneously remains an open research question.

Another challenge in evaluating projections is the need for metrics that can model human perception of class separability and other perceptual factors, especially in supervised multidimensional projection settings \cite{Grac}. Existing metrics, such as Distance Consistency (DSC), Dunn's index, LDA's objective, and the newly proposed GONG and KNNG metrics, have shown promise in this regard, but their computational efficiency and ability to capture density information in iterative projection processes still need to be improved \cite{DSC, Dunn, Caco, Sedl, Aupe}.

To address the evaluation challenges and limitations of traditional quality metrics, our research proposes a personalized projection framework that combines multiple quality metrics with user preferences to generate projections that better align with users' criteria and perceptual abilities. This framework aims to overcome the trade-offs between local and global structure preservation and incorporate human perception factors into the evaluation process.

The personalized projection framework involves the following components:

\begin{enumerate}
\item Generation of multiple projections using various projection techniques (e.g., t-SNE, UMAP, LAMP) and parameter configurations.
\item Calculation of a set of quality metrics (e.g., stress, neighborhood preservation, silhouette score) for each projection.
\item Collection of user ratings for a subset of projections based on their preferences and criteria.
\item Learning of optimal weights for each user by minimizing the difference between user ratings and a composite quality metric that combines the individual quality metrics.
\item Application of the learned weights to other datasets and the identification of the best projections for each user.
\end{enumerate}

By incorporating user preferences into the evaluation process, the personalized projection framework has the potential to generate projections that better reflect users' perceptual abilities and criteria, thereby facilitating more effective data exploration and interpretation. Additionally, the framework can help researchers and practitioners navigate the increasingly complex landscape of visualization techniques and parameter choices, ultimately improving the quality and impact of multidimensional data visualizations.

\section{Personalized Projection Framework}

In this section, we present a personalized projection framework that enables users to explore multidimensional data in a way that aligns with their preferences and criteria. Our framework consists of three main components: a composite quality metric, an optimization of projection scale, and user-specific weighting of quality metrics.

\subsection{Composite quality metric}

The composite quality metric combines three visual quality metrics—stress, neighborhood preservation, and silhouette score—to provide a more comprehensive evaluation of projection quality. By considering these three metrics, the composite quality metric captures different aspects of the projection, such as global and local structure preservation and cluster separation.

The composite quality metric is defined as follows:

\begin{equation}
Q(P) = w_1+ w_2\cdot \textit{SC}(P) + w_3 \cdot \textit{Stress}(P) + w_4 \cdot \textit{NP}(P)   
\end{equation}

where $Q(P)$ represents the composite quality metric for projection $P$, and $w_1$, $w_2$, and $w_3$ are user-defined weights for the stress, neighborhood preservation, and silhouette score metrics, respectively. These weights can be adjusted by the user to prioritize different aspects of the projection according to their preferences.

\subsection{Optimization of projection scale}

The optimization of projection scale aims to find the best projection scale for a given dataset and user preferences. The projection scale is an important factor that affects the visual quality of the projection, as it determines how the high-dimensional data is represented in the low-dimensional space.

To find the optimal projection scale, we maximize the value of the composite quality metric $Q(P)$:

\[ \textit{OptimalScale}(D) = \arg\max_{s \in S} Q(P_s(D)) \]

where $D$ is the dataset, $P_s(D)$ is the projection with scale $s$, and $S$ is the set of candidate projection scales. The optimization process can be performed using gradient-based or derivative-free optimization algorithms, such as the Nelder-Mead method or particle swarm optimization.

\subsection{User-specific weighting of quality metrics}

The personalized projection framework allows users to define their own weighting of the quality metrics to create projections that align with their preferences and criteria. For example, a user interested in the preservation of local neighborhood structures may assign a higher weight to the neighborhood preservation metric, while a user focused on the separation of clusters may prioritize the silhouette score metric.

To incorporate user-specific preferences, we can design an experiment where users rate different projections according to their personal criteria. This generates two different labeled training sets, T1 and T2, one for each user. We then learn the set of weights, W1 and W2, for each training set using machine learning techniques, such as linear regression or support vector machines.

Once the weights are learned, they can be applied to other datasets to find the best projections for each user. By analyzing the resulting projections, we can assess whether the personalized projection framework is coherent with the different users' criteria and preferences. This personalized approach can help users explore and interpret multidimensional data more effectively, as the projections are tailored to their specific needs and interests.

\section{Experiment Design}

In this section, we outline the experimental design for evaluating our personalized projection framework. We detail the datasets used, the generation of projections, the collection of user ratings and training sets, and the learning of optimal weights.

\subsection{Datasets}

To evaluate the performance of our personalized projection framework, we use three different datasets: Iris, Wine, and Digits. These datasets represent a variety of data types and structures, allowing us to test the framework's adaptability and robustness.
\begin{enumerate}
    \item Iris: The Iris dataset consists of 150 samples from three species of Iris flowers, with four features measured for each sample. This dataset is widely used in machine learning and visualization research, and its relatively low dimensionality makes it suitable for our experiment. 
    \item Wine: The Wine dataset contains 178 samples from three different types of wine, with 13 chemical features measured for each sample. This dataset presents a more challenging visualization task due to its higher dimensionality and complex structure.
    \item     Digits: The Digits dataset consists of $8\times 8$ pixel images of handwritten digits from $0$ to $9,$ resulting in 64 features for each sample. This dataset provides a unique challenge in visualizing high-dimensional image data with non-linear structures.
\end{enumerate}

\subsection{Generation of projections}

For each dataset, we generate projections using three different multidimensional projection techniques: t-SNE, UMAP, and LAMP. We vary the parameters of each technique to explore a range of projection configurations.
\begin{enumerate}
    \item     t-SNE: We vary the perplexity parameter in the range of $5$ to $50,$ with increments of $5.$

    \item   UMAP: We vary the $n-$neighbors parameter using a percentage of the total data points, with values ranging from $5\%$ to $50\%$, with increments of $5.$

    \item   LAMP: We vary the percentage of control points used for the projection, with values ranging from $10\%$ to $100\%$, with increments of $10.$
\end{enumerate}

\begin{table}[h!]
\centering
\begin{tabular}{|l|c|c|}
\hline
\textbf{Technique} & \textbf{Parameter} & \textbf{Configuration} \\ \hline
t-SNE & Perplexity & 5, 10, ..., 50 \\ \hline
UMAP & $n$-neighbors & 10\%, 20\%, ..., 100\% \\ \hline
LAMP & Control points & 10\%, 20\%, ..., 100\% \\ \hline
\end{tabular}
\caption{Parameter configurations used for generating projections with t-SNE, UMAP, and LAMP.}
\label{tab:parameter-configurations}
\end{table}

\subsection{User ratings and training sets}

We design an experiment in which two users, U1 and U2, with different projection preferences, rate the generated projections according to their personal criteria. This process generates two different labeled training sets, T1 and T2, one for each user.

To ensure that the user ratings are based on their visual perception and not on their knowledge of the projection techniques, we provide them with guidelines for assessing the quality of the projections based on visual cues, such as the preservation of local and global structures, separation of clusters, and overall clarity.

Table~\ref{tab:training_sets} shows the distribution of the training sets. There are 120 projections, 40 for each method, and each method has 10 projections of the 4 different datasets with different parameter configurations.

\begin{table}[ht]
\centering
\caption{Distribution of the training sets}
\label{tab:training_sets}
\begin{tabular}{llrrr}
\hline
Method & Dataset & User 1 Ratings & User 2 Ratings & Total \\
\hline
t-SNE & Iris & 10 & 10 & 20 \\
t-SNE & Wine & 10 & 10 & 20 \\
t-SNE & Digits & 10 & 10 & 20 \\
t-SNE & Breast Cancer & 10 & 10 & 20 \\
UMAP & Iris & 10 & 10 & 20 \\
UMAP & Wine & 10 & 10 & 20 \\
UMAP & Digits & 10 & 10 & 20 \\
UMAP & Breast Cancer & 10 & 10 & 20 \\
LAMP & Iris & 10 & 10 & 20 \\
LAMP & Wine & 10 & 10 & 20 \\
LAMP & Digits & 10 & 10 & 20 \\
LAMP & Breast Cancer & 10 & 10 & 20 \\
\hline
\textbf{Total} & & \textbf{120} & \textbf{120} & \textbf{240} \\
\hline
\end{tabular}
\end{table}

For the training and evaluation of the machine learning models, we use 80\% of the data as the training set and reserve 20\% as the test set. The training set is used to train the models and perform hyperparameter tuning, while the test set is used to evaluate the model's performance on unseen data.

\begin{figure*}[!t]
\centering
\subfloat[]{\includegraphics[width=2in]{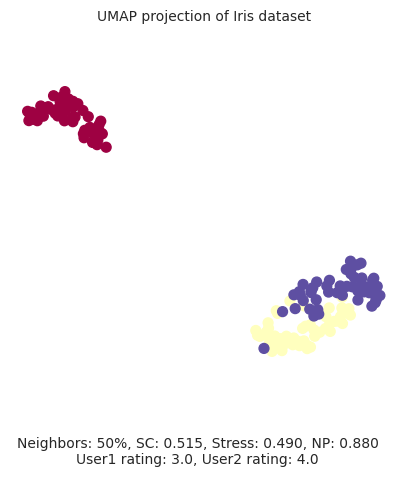}%
\label{fig:umap1}}
\hfil
\subfloat[]{\includegraphics[width=2in]{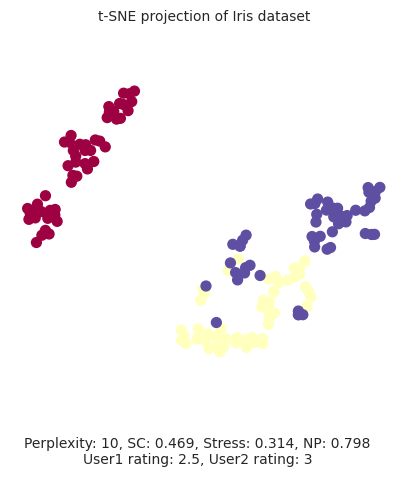}%
\label{fig:tsne1}}
\hfil
\subfloat[]{\includegraphics[width=2in]{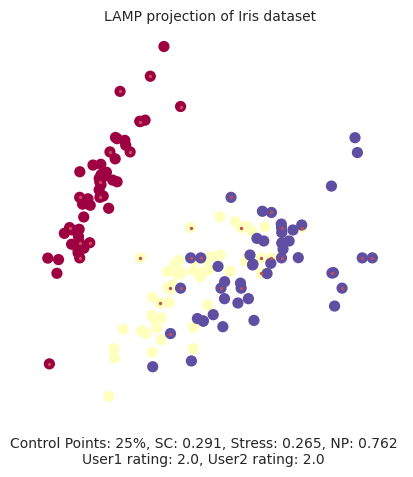}%
\label{fig:lamp1}}
\caption{Projections of the Iris dataset using UMAP, t-SNE, and LAMP.}
\label{fig:proj_rating}
\end{figure*}

\subsection{Learning Optimal Weights}

Once the user ratings and training sets have been collected, we learn the optimal weights for each user by fitting a model to their respective training sets. The model aims to predict the user's ratings based on the composite quality metric, which combines stress, neighborhood preservation, and silhouette score. We used linear regression, ridge regression, and Lasso regression models, along with cross-validation, to find the best model for each user.

\begin{table}[h!]
\centering
\begin{tabularx}{\linewidth}{|X|X|X|X|X|X|}
\hline
\textbf{User} & \textbf{Bias Weight} & \textbf{SC Weight} & \textbf{Stress Weight} & \textbf{Neighborhood Preservation Weight} & \textbf{Best Model} \\ \hline
U1 & 1.820 & 2.993 & 0.314 & 0.086 & Ridge Regression \\ \hline
U2 & 2.230 & 2.433 & -0.089 & 0.812 & Ridge Regression \\ \hline
\end{tabularx}
\caption{Learned weights for users U1 and U2, the corresponding quality metrics, and the best regression model selected for each user.}
\label{tab:learned-weights}
\end{table}

By following this experiment design, we can evaluate the effectiveness of the personalized projection framework in generating projections that align with user preferences and criteria, as well as assess its adaptability and robustness across different datasets and projection techniques. Furthermore, the use of different regression models and cross-validation ensures that the learned weights are reliable and result in the best possible predictions of user ratings.

The user-specific metrics derived from ridge regression for the optimal projection evaluation are as follows:

\begin{equation}
\text{User 1 Rating} = 1.820 + 2.993 \cdot \text{SC} + 0.314 \cdot \text{Stress} + 0.086 \cdot \text{NP}
\end{equation}

\begin{equation}
\text{User 2 Rating} = 2.230 + 2.433 \cdot \text{SC} + (-0.089) \cdot \text{Stress} + 0.812 \cdot \text{NP}
\end{equation}

\begin{figure*}[ht]
\centering
\subfloat[User 1.]{\includegraphics[width=2.5in]{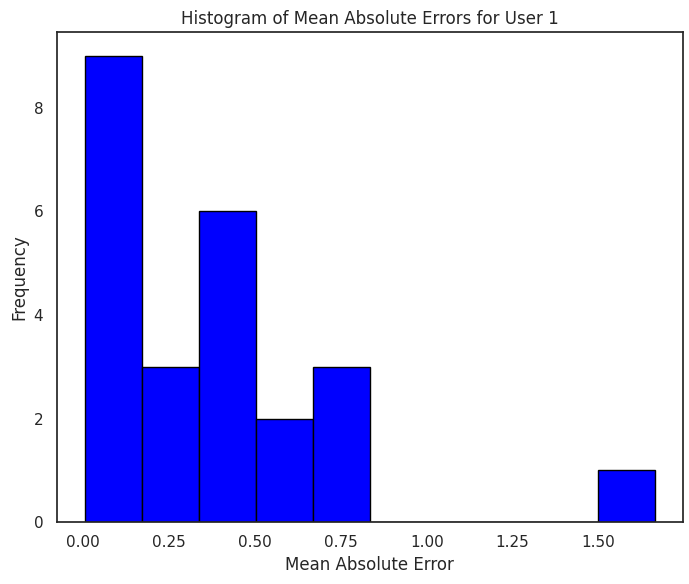}%
\label{fig:hist1}}
\hfil
\subfloat[User 2.]{\includegraphics[width=2.5in]{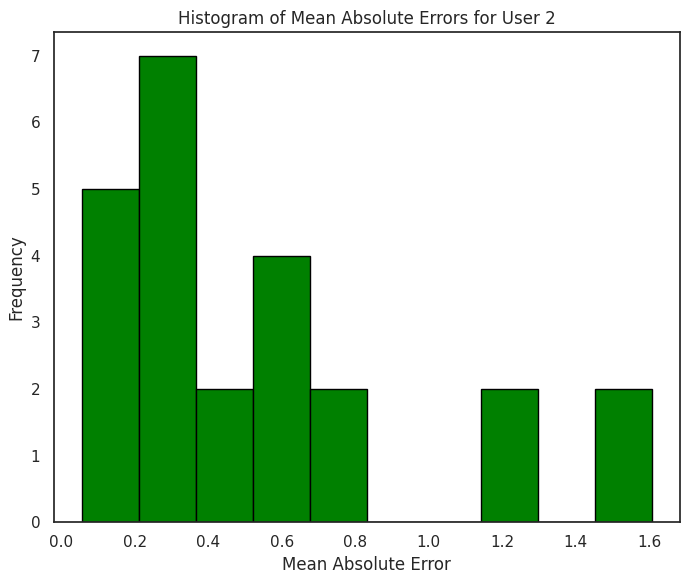}%
\label{fig:hist2}}
\caption{Histograms showing the distribution of mean absolute errors for the predicted user ratings using ridge regression.}
\end{figure*}

The histogram (Figure~\ref{fig:hist1}
) shows that the majority of the predictions for User 1 have a relatively low MAE, with the highest frequency of errors falling within the range of 0.0 to 0.25. There are fewer instances of higher errors, with a notable drop in frequency as the error increases. 

In contrast, the histogram for User 2 (Figure~\ref{fig:hist2}) exhibits a slightly different pattern. While there is still a high frequency of low MAE values between 0.2 and 0.4, there is also a noticeable presence of errors in the range of 0.6 to 1.6.
\section{Results and Evaluation}

In this section, we present the results of our personalized projection framework, including the generated projections for users U1 and U2.

\subsection{Projections for user U1}

\begin{figure*}[!t]
\centering
\subfloat[Iris dataset using t-SNE with a metric of 3.62]{\includegraphics[width=1.8in]{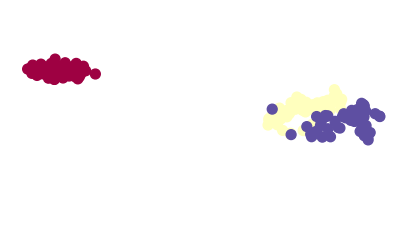}%
\label{fig_tsne_iris_U1}}
\hfil
\subfloat[Iris dataset using UMAP with a metric of 3.71]{\includegraphics[width=1.8in]{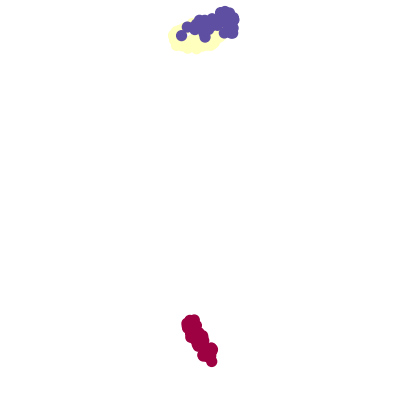}%
\label{fig_umap_iris_U1}}
\hfil
\subfloat[Iris dataset using LAMP with a metric of 3.1]{\includegraphics[width=1.8in]{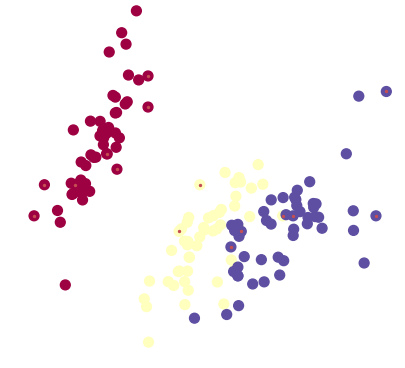}%
\label{fig_lamp_iris_U1}}
\hfil
\subfloat[Wine dataset using t-SNE with a metric of 3.66]{\includegraphics[width=1.8in]{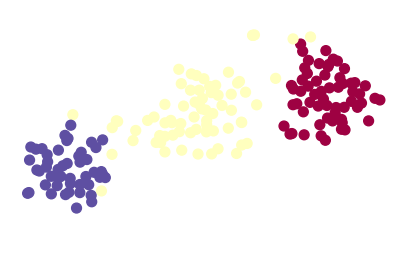}%
\label{fig_tsne_wine_U1}}
\hfil
\subfloat[Wine dataset using UMAP with a metric of 3.7]{\includegraphics[width=1.8in]{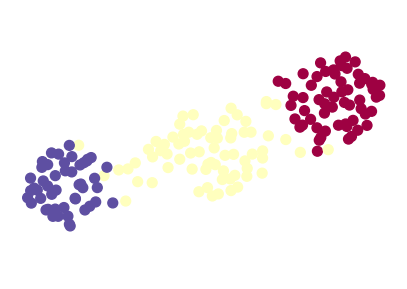}%
\label{fig_umap_wine_U1}}
\hfil
\subfloat[Wine dataset using LAMP with a metric of 3.25]{\includegraphics[width=1.8in]{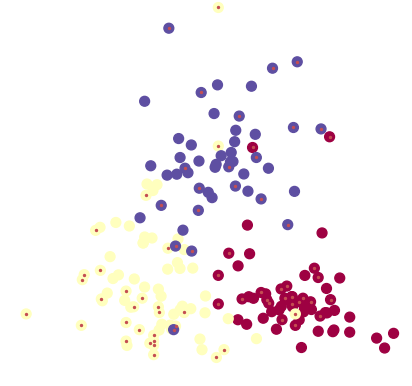}%
\label{fig_lamp_wine_U1}}
\hfil
\subfloat[Digits dataset using t-SNE with a metric of 3.6]{\includegraphics[width=1.8in]{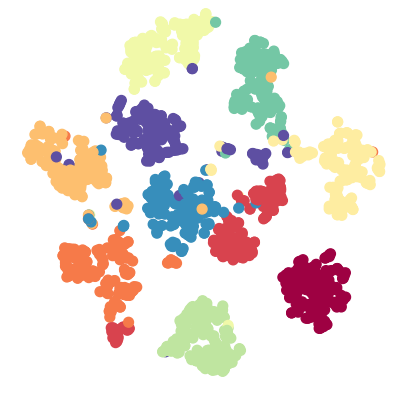}%
\label{fig_tsne_digits_U1}}
\hfil
\subfloat[Digits dataset using UMAP with a metric of 3.55]{\includegraphics[width=1.8in]{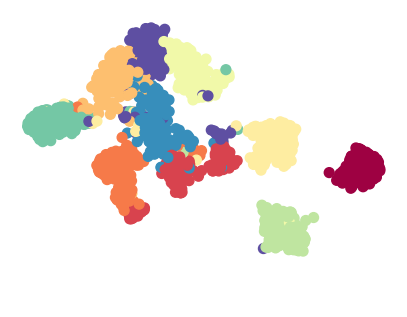}%
\label{fig_umap_digits_U1}}
\hfil
\subfloat[Digits dataset using LAMP with a metric of 1.61]{\includegraphics[width=1.8in]{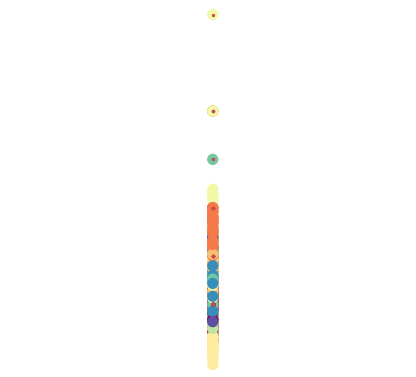}%
\label{fig_lamp_digits_U1}}
\hfil
\subfloat[Breast Cancer dataset using t-SNE with a metric of 3.58]{\includegraphics[width=1.8in]{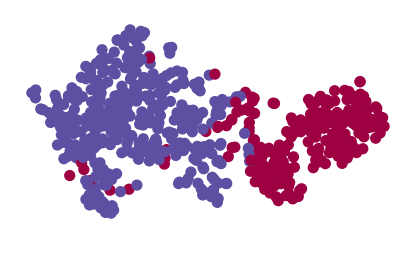}%
\label{fig_tsne_cancer_U1}}
\hfil
\subfloat[Breast Cancer dataset using UMAP with a metric of 3.58]{\includegraphics[width=1.8in]{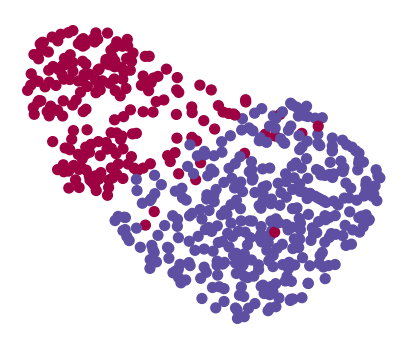}%
\label{fig_umap_cancer_U1}}
\hfil
\subfloat[Breast Cancer dataset using LAMP with a metric of 3.35]{\includegraphics[width=1.8in]{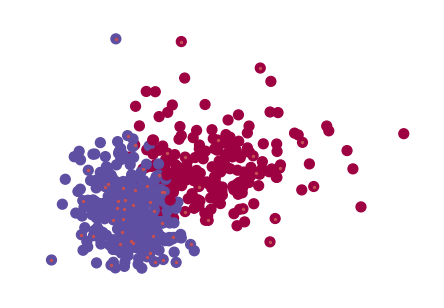}%
\label{fig_lamp_cancer_U1}}
\caption{Visualizations of the best projections for User 1 (U1) for each dataset (Iris, Wine, Digits, and Breast Cancer) using different projection techniques (t-SNE, UMAP, and LAMP) and their corresponding metrics.}
\label{fig:U1}
\end{figure*}

After applying the personalized projection framework for user U1, we generate projections tailored to their preferences and criteria. Figure \ref{fig:U1} and Table \ref{table:best_projections} should be inserted here, showing the best projections for U1 based on the composite quality metric and their respective parameter configurations.

The visualizations and parameter configurations presented in Figure \ref{fig:U1} and Table \ref{table:best_projections} demonstrate the effectiveness of the personalized projection framework in generating projections that align with user U1's preferences.

\begin{table}[htbp]
  \centering
  \caption{Parameter configurations of the best projections for user U1 for each dataset and projection technique.}
  \label{table:best_projections}
  \begin{tabularx}{\linewidth}{XXXX}
    \toprule
    Dataset       & t-SNE (Perplexity)        & UMAP (Neighbors)         & LAMP (Control points)         \\
    \midrule
    Iris          & \textit{$35$} & \textit{$70$} & \textit{$10$} \\
    Wine          & \textit{$30$} & \textit{$30$} & \textit{$40$} \\
    Digits        & \textit{$5$} & \textit{$10$} & \textit{$10$} \\
    Breast Cancer & \textit{$25$} & \textit{$80$} & \textit{$10$} \\
    \bottomrule
  \end{tabularx}
\end{table}

\subsection{Projections for user U2}

\begin{figure*}[!t]
\centering
\subfloat[Iris dataset using t-SNE with a metric of 4.2]{\includegraphics[width=1.8in]{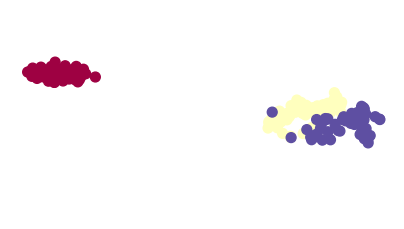}%
\label{fig_tsne_iris_U2}}
\hfil
\subfloat[Iris dataset using UMAP with a metric of 4.27]{\includegraphics[width=1.8in]{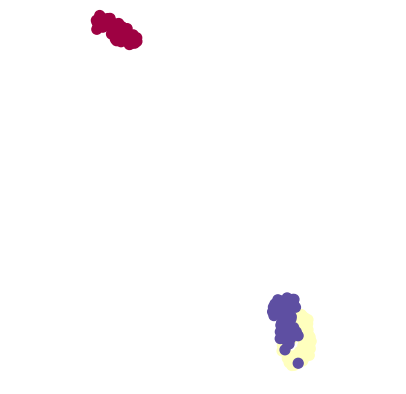}%
\label{fig_umap_iris_U2}}
\hfil
\subfloat[Iris dataset using LAMP with a metric of 3.82]{\includegraphics[width=1.8in]{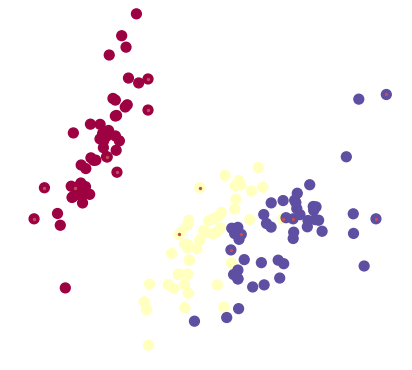}%
\label{fig_lamp_iris_U2}}
\hfil
\subfloat[Wine dataset using t-SNE with a metric of 4.21]{\includegraphics[width=1.8in]{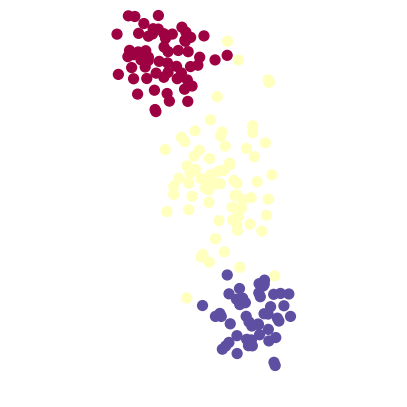}%
\label{fig_tsne_wine_U2}}
\hfil
\subfloat[Wine dataset using UMAP with a metric of 4.27]{\includegraphics[width=1.8in]{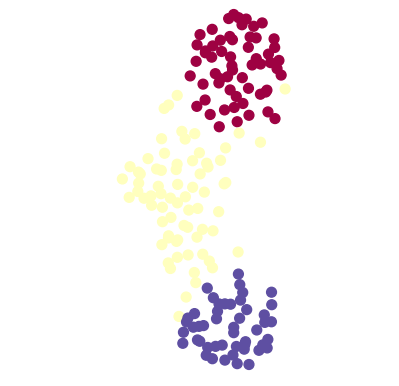}%
\label{fig_umap_wine_U2}}
\hfil
\subfloat[Wine dataset using LAMP with a metric of 3.7]{\includegraphics[width=1.8in]{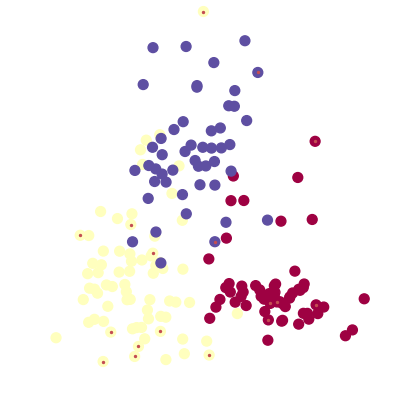}%
\label{fig_lamp_wine_U2}}
\hfil
\subfloat[Digits dataset using t-SNE with a metric of 3.48]{\includegraphics[width=1.8in]{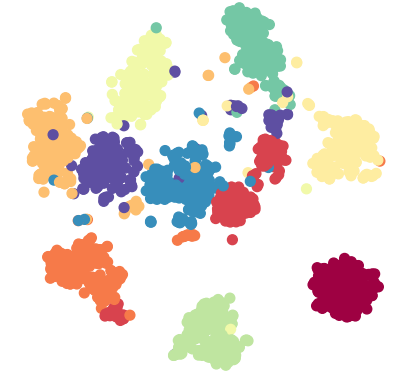}%
\label{fig_tsne_digits_U2}}
\hfil
\subfloat[Digits dataset using UMAP with a metric of 3.47]{\includegraphics[width=1.9in]{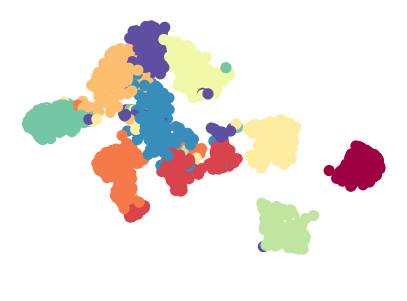}%
\label{fig_umap_digits_U2}}
\hfil
\subfloat[Digits dataset using LAMP with a metric of 1.77]{\includegraphics[width=1.8in]{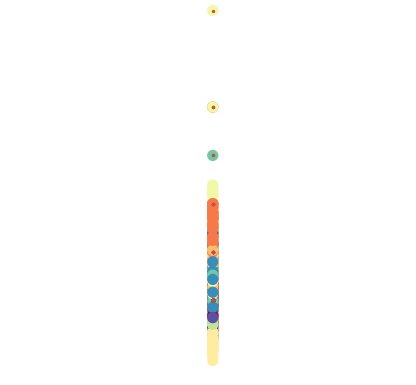}%
\label{fig_lamp_digits_U2}}
\hfil
\subfloat[Breast Cancer dataset using t-SNE with a metric of 3.84]{\includegraphics[width=1.8in]{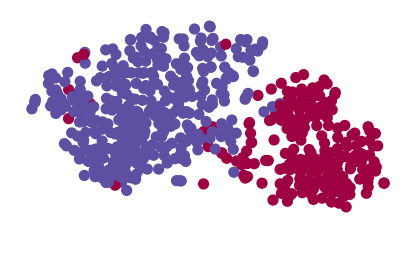}%
\label{fig_tsne_cancer_U2}}
\hfil
\subfloat[Breast Cancer dataset using UMAP with a metric of 4.12]{\includegraphics[width=1.8in]{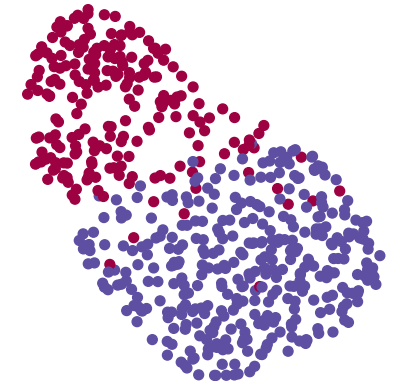}%
\label{fig_umap_cancer_U2}}
\hfil
\subfloat[Breast Cancer dataset using LAMP with a metric of 3.61]{\includegraphics[width=1.8in]{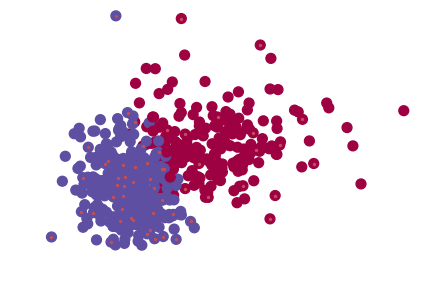}%
\label{fig_lamp_cancer_U2}}
\caption{Visualizations of the best projections for User 2 (U2) for each dataset (Iris, Wine, Digits, and Breast Cancer) using different projection techniques (t-SNE, UMAP, and LAMP) and their corresponding metrics.}
\label{fig:U2}
\end{figure*}

Similarly, we generate personalized projections for user U2 by applying the framework to their preferences and criteria. Figure \ref{fig:U2} and Table \ref{table:best_projections_U2} should be inserted here, illustrating the best projections for U2 and their corresponding parameter configurations.

The results in Figure \ref{fig:U2} and Table \ref{table:best_projections_U2} showcase the adaptability of the personalized projection framework to accommodate different user preferences and generate projections that meet user U2's specific criteria.

\begin{table}[htbp]
  \centering
  \caption{Parameter configurations of the best projections for user U2 for each dataset and projection technique.}
  \label{table:best_projections_U2}
  \begin{tabularx}{\linewidth}{XXXX}
    \toprule
    Dataset       & t-SNE (Perplexity)        & UMAP (Neighbors)         & LAMP (Control points)         \\
    \midrule
    Iris          & \textit{$35$} & \textit{$100$} & \textit{$10$} \\
    Wine          & \textit{$50$} & \textit{$100$} & \textit{$10$} \\
    Digits        & \textit{$50$} & \textit{$10$} & \textit{$10$} \\
    Breast Cancer & \textit{$45$} & \textit{$100$} & \textit{$10$} \\
    \bottomrule
  \end{tabularx}
\end{table}

\section{Discussion}

The results of our personalized projection framework demonstrate its effectiveness in generating high-quality visualizations tailored to individual user preferences. By incorporating user-specific quality metrics, our approach addresses the limitations of generic metrics that fail to capture diverse user criteria and perceptual differences.

Our study highlights the significance of personalized metrics in enhancing the interpretability and usability of multidimensional projections. The distinct weight configurations learned for users U1 and U2 reflect their unique preferences, which were effectively captured by the ridge regression model. For example, user U1's preference for higher silhouette scores and stress is evident in the weights assigned, while user U2's emphasis on neighborhood preservation is reflected in their weight configuration. These differences underscore the importance of accommodating individual user preferences in visualization tasks.

The ridge regression model proved to be the most effective in capturing user preferences and predicting user ratings, as indicated by its superior performance in cross-validation and test RMSE metrics. The stability of ridge regression in the presence of multicollinearity and its ability to shrink less significant coefficients contributed to its robustness in our framework. This robustness is particularly important when dealing with complex, high-dimensional data, where relationships between variables can be intricate and interdependent.

The experiment design, which involved generating and rating projections using t-SNE, UMAP, and LAMP, provided valuable insights into the adaptability and effectiveness of different projection techniques. The personalized approach allowed us to identify the optimal parameter configurations for each user, resulting in projections that better align with their criteria. Additionally, the use of different datasets (Iris, Wine, Digits, and Breast Cancer) demonstrated the framework's versatility and robustness across diverse data types and structures.

Despite its advantages, our framework has some limitations. The reliance on user ratings introduces subjectivity, which can vary significantly between users and contexts. Additionally, the computational complexity of generating multiple projections and training regression models can be a bottleneck for large datasets. Future work should explore methods to streamline the process and reduce computational overhead, potentially through more efficient algorithms or parallel processing techniques.

\section{Conclusions and Future Work}

In this study, we proposed a new framework for personalized multidimensional projections that tailors visualizations based on user-specific quality metrics. By combining stress, neighborhood preservation, and silhouette score into a composite metric and optimizing projection parameters accordingly, our approach enhances the interpretability and usability of high-dimensional data visualizations.

Our key contributions include the introduction of a composite quality metric, the optimization of projection scales based on user preferences, and the application of machine learning models to learn optimal weights for individual users. The effectiveness of our framework was demonstrated through experiments involving two users with distinct preferences, showcasing its adaptability and robustness across different datasets and projection techniques.

Future research should focus on several key areas. Scalability is crucial, and developing more efficient algorithms and computational techniques to handle larger datasets will reduce the time required for generating and evaluating projections. Implementing automated methods for collecting user feedback, potentially through interactive visual analytics tools, will allow users to adjust weights and parameters in real-time. Extending the framework to incorporate additional multidimensional projection techniques and quality metrics will further broaden its applicability and effectiveness. Conducting comprehensive user studies will validate the framework's effectiveness in real-world scenarios and gather insights into user behavior and preferences. Exploring methods to improve the interpretability of projections, such as incorporating domain-specific knowledge or visual aids, will help users better understand the relationships between high-dimensional data points.

The personalized projection framework represents a significant advancement in the field of high-dimensional data visualization. By prioritizing user preferences and optimizing projection techniques accordingly, we enable more effective data exploration and decision-making. This user-centric approach has the potential to drive further innovations in visualization techniques and improve the accessibility and utility of complex data for a wide range of users.

\section*{Acknowledgments}
The author gratefully acknowledges the financial support of the Science Foundation Ireland (SFI) under Grant Number SFI 18/CRT/6049.

\end{document}